\documentclass[letterpaper, 10 pt, conference]{ieeeconf}  % Comment this line out if you need a4paper

\IEEEoverridecommandlockouts                              % This command is only needed if 
\usepackage{cite}
\usepackage{amsmath,amssymb,amsfonts}
\allowdisplaybreaks[4]
\usepackage{algorithmic}
\usepackage{graphicx}
\usepackage{textcomp}
\usepackage{xcolor}    
\usepackage{url}   
\usepackage{epsfig}
\usepackage[T1]{fontenc}
\newtheorem{remark}{Remark}
\title{\LARGE \bf
Energy-Grade Double Pricing Rule in the Heating Market 
}

\author{Xinyi Yi \qquad Ye Guo \qquad Hongbin Sun% <-this % stops a space
\thanks{X. Yi and Y. Guo are with the Tsinghua-Berkeley Shenzhen Institute, Shenzhen, Guangdong 518055, China. H. Sun is with the Tsinghua University, Beijing 100084, China. This work is supported in part by the National Science Foundation of China under Grant 51977115.} % <-this % stops a space
\thanks{Corresponding author: Ye Guo, e-mail: {\tt\small guo-ye@sz.tsinghua.edu.cn}.}%
}

\begin{document}

\maketitle
\thispagestyle{empty}
\pagestyle{empty}

%%%%%%%%%%%%%%%%%%%%%%%%%%%%%%%%%%%%%%%%%%%%%%%%%%%%%%%%%%%%%%%%%%%%%%%%%%%%%%%%
\begin{abstract}

The problem of heat system pricing is considered. A direct extension of locational marginal prices (LMP) in electricity markets to heat systems may lead to revenue inadequate issues. The underlying reason for such a problem is that, unlike electric power, heat has different grades and cannot be considered as homogenized commodity. Accordingly, an energy-grade double pricing rule is proposed in this paper. Heat energy and grade prices are explained as the shadow prices related to the nodal heat balance constraints and temperature requirements constraints at the optimal solution. The resulting merchandise surplus at each dispatch interval can be decomposed into several explainable parts, namely, congestion rent, impact from the last period, and impact from the upcoming period. And the total merchandise surplus over all dispatch intervals can be decomposed into several non-negative interpretable parts, including congestion rent and impact from the initial state,  thus guaranteeing the revenue adequacy for the heat system operator. Simulations verify the effectiveness of the proposed mechanism.

\end{abstract}

%%%%%%%%%%%%%%%%%%%%%%%%%%%%%%%%%%%%%%%%%%%%%%%%%%%%%%%%%%%%%%%%%%%%%%%%%%%%%%%%
\section{INTRODUCTION}
 In the integrated energy system, different energy carriers can be coordinated to improve fuel efficiency and alleviate environmental pollution. For example, in the combined heat and power (CHP) system, fuel efficiency can be improved by 50\%, and carbon emission can be reduced by 13-18\% compared to single out-put generators because of the waste heat recycling and the complementarity of electric power systems and heat systems\cite{2018Integrated}. Therefore, a better-coordinated energy system, including sectors of electric power, heating, and cooling, has been extensively developed. In China, CHP units account for about
 40\% of installed thermal power by 2020 \cite{chinatest2}. And CHP units produced 20.1 TWh of electricity in the UK in 2016 \cite{Data}. However, the integration of the heat systems and electric power systems brings about the coupling of the two systems' operation. Their generation cost is also coupled with the CHP units' joint cost. Thus price regime in the CHP system needs to be further studied.
 
Whereas, compared to electricity markets, the pricing of heat is still an open question\cite{8c35ba4d790b4916b928bb9932a4586d}. The true-cost pricing principle is used in the regulated market like Russia \cite{2015A}, where prices and delivery conditions are supervised by national independent authorities. A cost-plus method was presented in \cite{2012Modernizing}. This pricing method has a high reliance on honest offers from generators and proper management of authorities. Some countries set heat prices based on a liberalized heat market. For example, in Sweden, the heating sector was opened to competition in 1996, and the marginal cost pricing mechanism is adopted\cite{iea}. Since the marginal cost pricing mechanism uses the cost of an alternative unit of heat supply in the heat system, which increases the transparency of price formulation, it becomes the mainstream of the research on heat pricing. The heat marginal energy price is proposed in a wholesale heat market in \cite{2019The}, and \cite{2019A} develops a dynamic pricing model based on the heat's levelized cost. In light of LMP's extensive adoption in the electricity market, paper \cite{2019Energy} optimizes the heat and electricity simultaneously and extends
the concept of LMP in electricity markets to heat pricing. Heat is priced as the shadow prices related to nodal heat energy balance at the optimal solution. Paper \cite{2019Generalized} further explains the meaning of heat energy marginal prices. The distribution law of the heat energy marginal prices is analyzed in \cite{2018Stackelberg}. In addition, to make full use of the CHP system's flexibility, the heat energy marginal price formulated with the participation of integrated demand response \cite{6693775} and flexible ramping product \cite{2021Network} is studied. 

However, most existing works on heat pricing focus on the costs of energy. Considering the important assumption of LMP that the electric power is homogeneous, it leads to locational uniform prices. For heat energy, however, different temperatures correspond to different qualities, or grades, of the heat energy\cite{9133375}. Paper \cite{2011Low} discusses the influence of the heat carrier's temperature on the production efficiency of heat generators. And paper \cite{6895131} presents the effect of the heat carrier's temperature on the consumers' comfort.

The main contributions of the paper are two-fold: (i) an energy-grade double pricing mechanism is proposed. After introducing optimization models used in heat system dispatch, we show by a toy example that a direct extension of LMP based on the model in heat systems may lead to problems of inadequate revenue. Subsequently, we develop the rule of energy-grade double pricing; (ii) We decompose the associated merchandise surplus of the heat system operator under the proposed rule into several interpretable components. The pricing rule explains how the heat energy and grade requirements affect the system's operation cost.

\section{Problem modeling}
In this section, we describe the operation model and the optimization model of the heat system respectively.  
\subsection{Operation model}
A heat system consists of heat exchangers, a supply network, and a return network as shown in Fig. \ref{fig 1}. The networks are composed of pipelines. Each heat exchanger has a pair of locations in the supply and return networks.  We use $ \tau$ and $T$ to denote the pipeline outlet temperatures and locational temperatures in the following parts.

\subsubsection{Heat node} 
\begin{figure}[htbp]
	\vspace{-2.2cm}
	\hspace{-16.0cm}
	\includegraphics[width=9.5in]{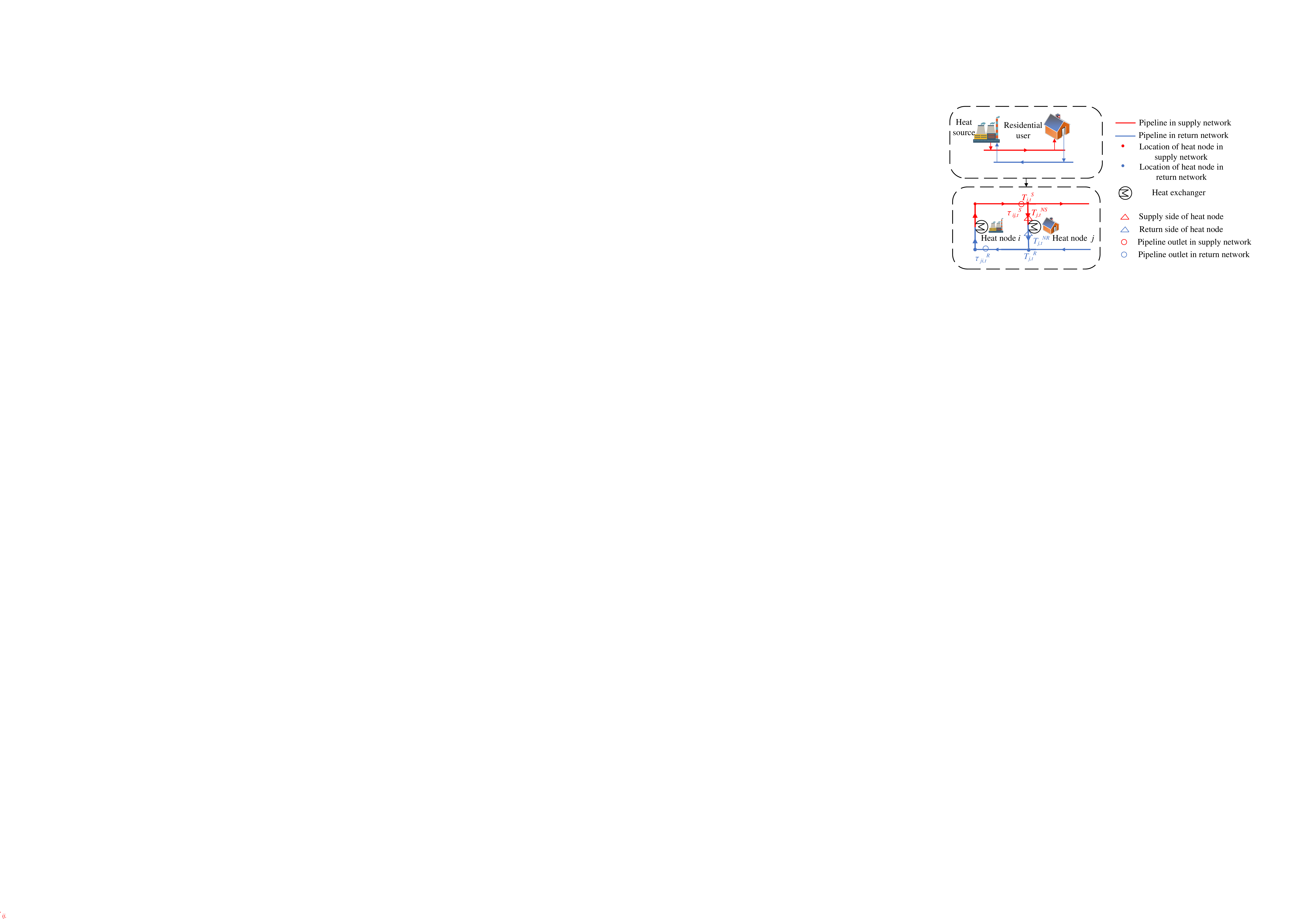}
	\vspace{-12.3cm}
	\caption{The general structure of a heating system.}
	\label{fig 1}
\end{figure}

Heat exchangers are represented by heat nodes in Fig \ref{fig 1} because the quantity of heat nodes' production and consumption is determined by the temperature difference between exchangers' supply and return sides. 
	\vspace{-0.4cm}

The nodal heat balance is:
\begin{subequations}
\begin{align}
    &G_{i,t}=cM_{i,t}(T_{i,t}^{NS}-T_{i,t}^{NR}), i\in \Psi_{S},\\ 
    &D_{i,t}=cM_{i,t}(T_{i,t}^{NS}-T_{i,t}^{NR}),i\in \Psi_{L},
    \end{align}
\end{subequations}
where $T_{i,t}^{NS}/T_{i,t}^{NR}$ represent the heat node $i$'s supply/return side temperature at period $t$. $G_{i,t}/D_{i,t}$ represent heat node $i$'s production/load power. $\Psi_{S/L}$ indicates the set of source/load nodes. $M_{i,t}$ is the mass flow through the heat exchanger of node $i$ at period $t$. $c$ is the specific heat capacity of the water.

For load nodes, water in the supply network injects into their heat exchangers, and is mixed in their return-side location. For source nodes, water with low temperature in the return network injects into their heat exchangers, and is mixed in their supply-side location. Thus the load/source nodal heat balances can be presented using their return/supply locational temperature mixing respectively:
\begin{subequations}\label{nodel balance}
	\begin{align}
		G_{hi,t}-D_{hi,t}=&c[(M_{i,t}+\sum_{r\in\Psi(i)}m_{r,t})T_{1i,t}\\\notag
		&-M_{i,t}T_{2i,t}-\sum_{r\in\Psi(i)}m_{r,t}\tau_{r,t}],\\
		T_{1i,t}=&\left\{
		\begin{aligned}
			T_{i,t}^R & , & \forall i\in\Psi_{L}, \\
			T_{i,t}^S & , & \forall i\in\Psi_{S},\\
		\end{aligned}
		\right.\\
		T_{2i,t}=&\left\{
		\begin{aligned}
			T_{i,t}^S& , &  \forall i\in\Psi_{L}, \\
			T_{i,t}^R & , & \forall i\in\Psi_{S},\\
		\end{aligned}
		\right.\\
		\Psi(i)=&\left\{
		\begin{aligned}
			\Psi_R(i)& , &  \forall i\in\Psi_{L}, \\
			\Psi_S(i) & , & \forall i\in\Psi_{S},\\
		\end{aligned}
		\right.
	\end{align}
\end{subequations}
where $\Psi_{R}(i)$ and $\Psi_{S}(i)$ indicate the set of pipelines injecting into the location of node $i$ in the return network and supply network respectively. $m_{r,t}$ denotes the mass flow rate of pipeline $r$. $\tau_{r,t}$ denotes the outlet temperature of pipe $r$ at $t$. 

And load/source node $i$'s supply/return locational temperature mixing is presented as:
\begin{subequations}
\begin{align}
		0&=(\sum_{r\in \Psi_{S}(i)}m_{r,t})T_{i,t}^S-\sum_{r\in \Psi_{S}(i)}m_{r,t}\tau_{r,t}, i\in\Psi_{L},\\
		0&=(\sum_{r\in \Psi_{R}(i)}m_{r,t})T_{i,t}^R-\sum_{r\in \Psi_{R}(i)}m_{r,t}\tau_{r,t}, i\in\Psi_{S}.
		\end{align}\label{temperature mixing}
\end{subequations}
Considering the heat transfer process in pipelines, pipeline $r$'s outlet temperature $\tau_{r,t}$ can be calculated by its inlet nodal locational temperatures as described briefly in the following, please refer to \cite{7243359} for more details.

\subsubsection{Heating pipeline} 
Supply and return networks consist of pipelines connecting different nodes. Under the assumption of a day-ahead dispatch interval, the outlet temperature $\tau_{ji,t}^{S/R}$ of pipeline $ji$ in supply/return network at $t$ can be expressed by its inlet node temperature at $t$ and $t-1$\cite{2019Generalized}. When water flows through the pipeline, the temperature drops with distance. So the outlet temperature $\tau_{ji,t}^{S/R}$ is:
\begin{equation}\label{loss trans}
\begin{aligned}
	\tau_{ji,t}^{S/R}&=[(1-\psi_{ji}^{S/R})T_{j,t}+\psi_{ji}^{S/R}T_{j,(t-1)}\\
	&-T_{ai,t}]e^{-v*L_{ji}^{S/R}/cm_{ji}^{S/R}}
	+T_{ai,t}^{S/R},\\
	\end{aligned}
\end{equation}
where $\psi_{ji}^{S/R}=\rho S_{ji}^{S/R} L_{ji}^{S/R} /(m_{ji,t}^{S/R} \Delta T)$. $\rho$, $S_{ji}^{S/R}$,  $L_{ji}^{S/R}$, $m_{ji}^{S/R}$ and $\Delta T$  indicate the density of water, cross-section area, length, mass flow rate of pipeline from node $j$ to $i$ in supply/return networks and dispatch time interval respectively. The pipeline $ji$'s heat loss factor is $e^{-v*L_{ji}^{S/R}/cm_{ji}^{S/R}}$. $v$ and $T_{ai,t}^{S/R}$ indicate heat transfer coefficient per unit length of pipeline $ji$, and node $i$'s ambient temperature of supply/return locations respectively.$\tau_{r,t}$ with inlet node $j$ and outlet node $i$ in (\ref{nodel balance}-\ref{temperature mixing}) can be calculated as (\ref{loss trans}).
\subsubsection{Heating Network}
A $n$-node heat system can be represented by a graph consisting of $2n$ locations, $n$ exchangers, and $k$ pipelines. In this paper, we use the outset matrix $A_{a1}$ and the extremity matrix $A_{a2}$ to describe the connection in the graph \cite{2019Generalized}. Substituting (\ref{loss trans}) into (\ref{nodel balance})-(\ref{temperature mixing}) and listing (\ref{nodel balance})-(\ref{temperature mixing})  in sequence, the matrix form of (\ref{nodel balance})-(\ref{temperature mixing}) is:
\begin{equation}
	\boldsymbol{H_{t}=C_{1}T_{t}+C_{2}T_{t-1}+R_{t}}, \label{matrix e}
\end{equation}
where $\boldsymbol{T_{t}}$ is the vector of location temperatures at $t$. The element $H_{k,t}$ in the $k^{th}$ row of $\boldsymbol H_{t}$ is dependent on the constraint type of the corresponding row $k$ of the matrix. When the row $k$ corresponds to the node $i$'s heating balancing equation (\ref{nodel balance}), $H_{k,t}=G_{i,t}-D_{i,t}$. And when the row $k$ corresponds to the temperature mixing equation (\ref{temperature mixing}), $H_{k,t}=0$.  $\boldsymbol{C_{1}}$, $\boldsymbol{C_{2}}$ and $\boldsymbol{R_{t}}$ are parameters calculated as:
\begin{equation}
	\begin{aligned}
	    &\boldsymbol{C_{1}}=\boldsymbol{A_{a1}G_{c}A_{a1}^{T}-A_{a2}G_c DaA_{a1}^{T}},\\
		&\boldsymbol{C_{2}}=\boldsymbol{-A_{a2}G_{c}DbA_{a1}^{T}},\\
		&\boldsymbol{R_{t}}=\boldsymbol{A_{a2}G_{c}(D-I)A_{a1}^TT_{a,t}}.\\
		\end{aligned}
\end{equation}
$\boldsymbol D$, $\boldsymbol G_C$, $\boldsymbol a$ and $\boldsymbol b$ are calculated as:
\begin{equation}
	\begin{aligned}
		&\boldsymbol{D}=\boldsymbol{I}-v\boldsymbol{L}/c\boldsymbol{m}, \boldsymbol{G_{c}}=c\boldsymbol{m},\\
		&\boldsymbol a=\boldsymbol I-\rho \boldsymbol{S}\boldsymbol{L} /(\boldsymbol{m} \Delta T), \boldsymbol b=\rho \boldsymbol{S}\boldsymbol{L} /(\boldsymbol{m} \Delta T),\\
	\end{aligned}
\end{equation}
where $\boldsymbol{S}$, $\boldsymbol{L}$, $\boldsymbol{m}$ and $\boldsymbol{I}$ are diagonal matrices of pipelines' cross section area, length and mass flow rate, and 1, respectively.  $\boldsymbol{T_{a,t}}$ is the vector of ambient temperatures. The size of the matrices and vectors are defined in TABLE \ref{table 1}.
 	\vspace{-0.2cm}
 \begin{table}[ht]
	\footnotesize
	\centering
	\setlength{\abovecaptionskip}{-0.5pt}
	\setlength{\belowcaptionskip}{0pt}
	\caption{Size of the matrices and vectors}
	 \label{table 1}
	\begin{tabular}{cccc}
		\hline
		Matrix and vector&size\\
		\hline
		$\boldsymbol{C_{1}},\boldsymbol{C_{2}}$&2n*2n\\
		$\boldsymbol{T_{t}},\boldsymbol{T_{a,t}}\boldsymbol{H_{t}},\boldsymbol{R_{t}}$ &2n*1\\
     	$\boldsymbol{A_{a1}},\boldsymbol{A_{a2}}$ & 2n*(n+k)\\
		$\boldsymbol{S},\boldsymbol{L},\boldsymbol{m},\boldsymbol{I},\boldsymbol{G_c},\boldsymbol{D},\boldsymbol{a},\boldsymbol{b}$ & (n+k)*(n+k)\\
		\hline
	\end{tabular}
\end{table}
	\vspace{-0.8cm}

\subsection{Optimization model}
We assume that the heat load is inelastic and the mass flow rate is constant \cite{2019Generalized}. The optimal dispatch model of a heat system is formulated as:
\begin{equation}
	\begin{aligned}
		\min_{\boldsymbol{G_{t},T_{t}}} C_{H}(\boldsymbol{G_{t}})&=\min_{x}\sum_{t\in T}(\sum_{i\in \Psi_{CHP}}C_{i,t}^{CHP}+\sum_{i\in \Psi_{HB}}C_{i,t}^{HB}),\\
		x&={G_{i,t}^{CHP},G_{i,t}^{HB},T_{i,t}},
	\end{aligned}\label{of}
\end{equation}
$s.t.$
\begin{subequations}\label{cons}
\begin{align}
    &\boldsymbol{\lambda _{t}:H_{t}=C_{1}T_{t}+C_{2}T_{t-1}+R_{t}},\label{Heating balancingtemperature mixing constraint}\\
	&\boldsymbol{\underline{\gamma_{t}}, \overline{\gamma_{t}}:\underline {G}\le G_{t}\le\overline {G}},\label{Energy Source capacity constraint }\\
	&\boldsymbol{\mu_{t}: T_{t}\le T_{sa}},\label{Pipeline security constraints on temperature }\\
	&\boldsymbol{\beta_{t}:T_{t}\ge T_{Q}}.\label{Temperature requirements of heat sources and consumers }
	\end{align}
\end{subequations}
\subsubsection{Objective function}
There are two typical types of heat sources- heat boilers and CHP units. Their cost functions are:
\begin{equation}
	\begin{split}
		C_{i,t}^{HB}&=\mathcal{A}_{i}G_{i,t}^{HB}+\mathcal{B}_{i}{G_{i,t}^{HB}}^2, i\in \Psi_{HB}\\
		C_{i,t}^{CHP}&=\mathcal{C}_{i}G_{i,t}^{CHP}+\mathcal{D}_{i}{G_{i,t}^{CHP}}^2, i\in \Psi_{CHP}
	\end{split}
\end{equation}
where $\Psi_{HB}$ and $\Psi_{CHP}$ represent the set of heat boilers and CHP units. $\mathcal{A}_{i}$ and $\mathcal{B}_{i}$ are cost coefficients of heat boiler $i$, while $\mathcal{C}_{i}$ and $\mathcal{D}_{i}$ are cost coefficients of CHP unit $i$. The objective function minimizes the total cost of all generation units over the $T$-hour model horizon. 

\subsubsection{Constraints}
There are four kinds of constraints:
\begin{itemize}
	
	\item Heating balancing \& temperature mixing constraints (\ref{Heating balancingtemperature mixing constraint}), whose shadow prices are denoted by $\boldsymbol{\lambda _{t}}$. 
	\item Heat production constraints(\ref{Energy Source capacity constraint }), where $\boldsymbol{\underline {G}}$ and $\boldsymbol{\overline {G}}$ represent lower and upper bounds of heat sources' power output, whose shadow prices are denoted by $\boldsymbol{\underline{\gamma_{t}}}$ and $\boldsymbol{\overline{\gamma_{t}}}$. 
	\item Pipeline security constraints on temperature (\ref{Pipeline security constraints on temperature }), whose shadow prices are denoted by $\boldsymbol{\mu_{t}}$. It means that the temperature should not exceed its upper limit $\boldsymbol{T_{Sa}}$.
	\item  Temperature requirements of heat sources and consumers (\ref{Temperature requirements of heat sources and consumers }), whose shadow prices are denoted by $\boldsymbol{\beta_{t}}$. Heat consumers may have temperature requirements to ensure comfort. Heat sources may propose temperature requirements to enhance production efficiency.
\end{itemize}

\section{Energy-based pricing rule}
\subsection{Pricing and settlement}
We first implement a direct extension of the LMP in the electricity market to the pricing of heat. The Lagrangian of the optimal dispatch model (\ref{of}-\ref{cons}) is:
\begin{equation}
	\begin{split}
		L&=C_{H}+\sum_{t=1}^T[\boldsymbol{\lambda_{t}^T(C_{1}T_{t}+C_{2}T_{t-1}+R_{t}-H_{t})}\\
		&\boldsymbol{+\mu_{t}^T(T_{t}-T_{sa})+\beta_{t}^T(T_{Q}-T_{t})}\\
		&\boldsymbol{+\overline{\gamma_{t}}^T(G_{t}-\overline {G})+\underline{\gamma_{t}}^T(\underline {G}-G_{t})]}.
	\end{split}
\end{equation}

Similar to electricity markets, the LMP of heat node $i$ at time $t$ for heat is defined as the incremental cost to serve one unit additional load demand $D_{i,t}$ \cite{9281770}. When heat node $i$'s heat balance corresponds to the $k^{th}$ row in (\ref{matrix e}), the LMP of heat node $i$ at time $t$ is calculated as:
\begin{equation}
	p_{t}^i= \frac{\partial C_{H}^*}{\partial D_{i,t}}=\frac{\partial L_{CP}}{\partial H_{k,t}}=\lambda_{k,t}^*.
\end{equation}

Consequently, the payment of load at node $i$ is:
\begin{equation}
	\pi_{i}^L=\sum_{t=1}^T\pi_{i,t}^L=\sum_{t=1}^T p_{t}^i*D_{i,t}.
\end{equation}

Generator at node $i$ receives payment:
\begin{equation}
	\pi_{i}^S=\sum_{t=1}^T\pi_{i,t}^S=\sum_{t=1}^Tp_{t}^i*G_{i,t}^*.
\end{equation}

\subsection{Merchandise surplus}
The total merchandise surplus with these LMPs over T-period is calculated as:
\begin{equation}
	\mathcal{M}=\sum_{i\in \Psi_{L}}\pi_{i}^L-\sum_{i\in \Psi_{S}}\pi_{i}^S=\sum_{t=1}^T\sum_{i\in \Psi_{HN}}p_{t}^i*(D_{i,t}-G_{i,t}^*),
\end{equation}
where $\Psi_{HN}$ indicate the sets of heat nodes.

\begin{figure}[htbp]
	\vspace{-0.5cm}
	\hspace{2.4cm}
	\includegraphics[width=1.3in]{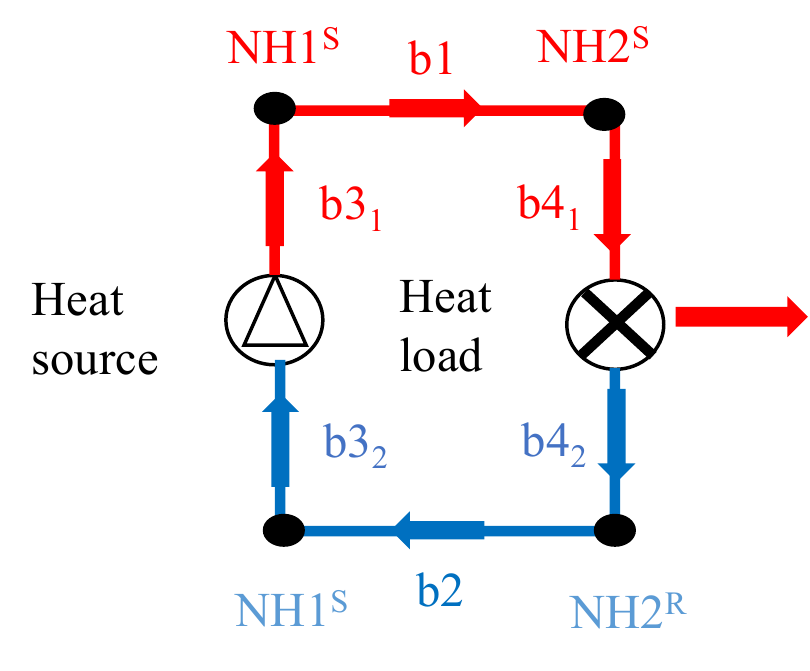}
	\vspace{-0.5cm}
	\caption{Heating system of the toy case.}
	\label{fig 3}
\end{figure}

Next, we illustrate LMPs with a toy heat system as shown in Fig. \ref{fig 3}, including a heat source node and a load node. For simplicity, we assume that the system has the same operation state at any time of the T-period and the ambient temperature keeps at -16$^{\circ}$C. The operation state and the solution to the dispatch model are shown in TABLE \ref{table 2} and \ref{table 3}, respectively. 
	\vspace{-0.3cm}

\begin{table}[ht]
	\footnotesize
	\centering
	\setlength{\abovecaptionskip}{-0.5pt}
	\setlength{\belowcaptionskip}{0pt}
	\caption{Operation state of the toy case}
		 \label{table 2}
	\begin{tabular}{cccc}
		\hline
		States &Value\\
		\hline
		Heat load ($MW$)&2\\
		Source type &CHP\\
		Cost coefficient & $\mathcal{C}=14.8$,$\mathcal{D}=0.0245$\\
		Energy source capacity ($MW$) & 4.0\\
		$v$ of $b1$ ($W/m^{\circ}C$) & 0.099\\
		$v$ of $b2$ ($W/m^{\circ}C$) & 0.099\\
		$L_{b1}$ (m)& 9000\\
		$L_{b2}$ (m)& 9000\\
		$b1$'s mass flow rate ($t/h$)& 1000\\
		$b2$'s mass flow rate ($t/h$)& 1000\\
		$b3$'s mass flow rate ($t/h$)& 1000\\
		$b4$'s mass flow rate ($t/h$)& 1000\\
		NH$1^S$ temperature requirement ($^{\circ}$C)&60-100\\
		NH$2^S$ temperature requirement ($^{\circ}$C)&60-100\\
		NH$1^R$ temperature requirement ($^{\circ}$C)&40-100\\
		NH$2^R$ temperature requirement ($^{\circ}$C)&30-100\\
		\hline
	\end{tabular}
\end{table}
	\vspace{-0.3cm}
	
From TABLE \ref{table 3}, we can see that the LMP of the source node is higher than that of the load node. In addition, the heat power production is greater than the power consumption due to network loss. Therefore, the merchandise surplus of the system operator is negative at any time period.

\vspace{-0.4cm}
\begin{table}[ht]
	\footnotesize
	\centering
	\setlength{\abovecaptionskip}{-0.5pt}
	\setlength{\belowcaptionskip}{0pt}
	\caption{Testing results}
		 \label{table 3}
	\begin{tabular}{cccc}
		\hline
		Results &Value\\
		\hline
		Generator's LMP ($\$/MWh$) &14.328\\
		Load's LMP ($\$/MWh$)& 14.317\\
		Merchandise surplus ($\$/h$)& -2.03\\
		Heat production ($MWh$)&2.14\\
		NH$1^S$ temperature ($^{\circ}$C)&60.06\\
		NH$2^S$ temperature ($^{\circ}$C)&60\\
		NH$1^R$ temperature ($^{\circ}$C)&58.23\\
		NH$2^R$ temperature ($^{\circ}$C)&58.29\\
		\hline
	\end{tabular}
	%\vspace{-0.2cm}
\end{table}

Next, we explain the smaller LMP of heat load. We can see that the temperature at NH$2^S$ reaches its lower bound in TABLE \ref{table 3}. So when the load increases by one unit, the temperature of NH$2^R$ will decrease while the temperature of NH$2^S$ will keep unchanged, which leads to less network loss in the return network and the same network loss in the supply network. So the loss component of the load's LMP is negative, and the load has a smaller LMP. Since lower bounds of nodes represent the nodal requirements for heat quality. This is the underlying reason for the negative merchandise surplus- heat is heterogeneous commodity which is different from the case of electricity. So we propose an energy-grade double pricing rule in the next section.

\section{Energy-grade double pricing rule}
In this section, we develop an energy-grade double pricing mechanism considering the different grade requirements of heat generators and consumers.

\subsection {Pricing and settlement}
For heat node $i$, it has quality requirements for both $T_{i,t}^S$ and $T_{i,t}^R$, which causes additional cost. The cost should be attributed to the requested nodes. In the energy-grade double pricing rule, we use heating grade price based on the marginal contribution of personalized requirements to the expected system total cost (\ref{of}) to reflect this. Based on the optimal dispatch model, according to the envelope theorem, the marginal grade price of heat node $i$ at $t$ is:
\begin{equation}
	{p_{g,t}^{Si}}=\frac{\partial C_H^*}{\partial T_{Qi,t}^S}=\frac{\partial L_{CP}}{\partial  T_{Qi,t}^S}=\beta_{i,t}^{S*},
\end{equation}
\begin{equation}
	{p_{g,t}^{Ri}}=\frac{\partial C_H^*}{\partial T_{Qi,t}^R}=\frac{\partial L_{CP}}{\partial  T_{Qi,t}^R}=\beta_{i,t}^{R*}.
\end{equation}
Note that the demand for heating quality is reflected in raising the water's temperature from the ambient temperature, in the proposed rule, we use the temperature difference between sources'/loads' personalized required temperatures and ambient temperature when considering the heat grade settlement and analyzing merchandise surplus.

The heat energy is settled at the LMP, which is the same as the energy-based pricing rule. The heat grade is settled at a marginal grade price. 
Thus the payment of load node $i$ is:
\begin{equation}
	\begin{split}
		\Upsilon_{i}^L&=\sum_{t=1}^T\Upsilon_{i,t}^L=\sum_{t=1}^T[(p_{t}^i*D_{i,t})+(p_{g,t}^{Si}*(T_{Qi,t}^S\\
		&-T_{ai,t})+p_{g,t}^{Ri}*(T_{Qi,t}^R-T_{ai,t}))], i\in \Psi_{L}.
	\end{split}
\end{equation}

Source node $i$ receives payment:
\begin{equation}
	\begin{split}
		\Upsilon_{i}^S&=\sum_{t=1}^T\Upsilon_{i,t}^S=\sum_{t=1}^T[(p_{t}^i*G_{i,t}^*)-(p_{g,t}^{Si}*(T_{Qi,t}^S\\
		&-T_{ai,t})+p_{g,t}^{Ri}*(T_{Qi,t}^R-T_{ai,t}))], i\in \Psi_{S}.
	\end{split}
\end{equation}

\subsection {Merchandise surplus}
The merchandise surplus of the system operator at time $t$ is calculated as:
\begin{subequations}\label{countt}
	\begin{align}
		\Upsilon_{t}&=\sum_{i\in \Psi_{L}}\Upsilon_{i,t}^L-\sum_{i\in \Psi_{S}}\Upsilon_{i,t}^S\\
		&=\boldsymbol{\mu_{t}^{*T}(T_{sa}-T_{a,t})}-\boldsymbol{\lambda_{t}^{*T}C_{2}(T_{t-1}^*}\\\notag
        &\boldsymbol{-T_{a,t})}+\boldsymbol{\lambda_{t+1}^{*T}C_{2}(T_{t}^*-T_{a,t})}\\
        &=CR_{t}+IL_{t}+IU_{t}.
	\end{align}
\end{subequations}		

For the detailed derivation of (\ref{countt}), please refer to appendix.
$\Upsilon_{t}$ can be decomposed into following three parts by their corresponding constraints: 
\subsubsection{Congestion rent $ CR_{t}$}	If the pipelines' security constraints on temperature (\ref{Pipeline security constraints on temperature }) are exceeded, the load of some nodes is unable to use the most economical unit in the heating system. In (\ref{Pipeline security constraints on temperature }), the pipelines' security constraints are all positive upper bounds, so $ CR_{t}$ are non-negative, like congestion rent in the electricity market.
\begin{equation}
	CR_{t}=\boldsymbol{\mu_{t}^{*T}(T_{sa}-T_{a,t})}\geq0.
\end{equation}
\subsubsection {Impact from the last period  $ IL_{t}$	}
Because of the thermal inertia, the temperature and heat of the previous period provide a base for the current operation, temperatures at $t$ are influenced by that at $t-1$. In particular, the impact from the last period is non-negative.
\begin{equation}
	IL_{t}=\boldsymbol{-\lambda_{t}^{*T}[C_{2}(T_{t-1}^*-T_{a,t})]}\geq0.
\end{equation}
\subsubsection {Impact from the upcoming period  $IU_{t}$	}
Similarly, temperatures at $t$ are influenced by that at $t+1$. It can be understood as an advance payment for future operation. So the impact from the upcoming period is non-positive.
\begin{equation}
	IU_{t}=\boldsymbol{\lambda_{t+1}^{*T}[C_{2}(T_{t}^*-T_{a,t})]}\le 0.
\end{equation}

\begin{remark}
The ramping constraints in the electricity market and the heat nodal balance constraints (9a) in the heat market are both interval-coupling constraints, but they influence the price formulation and pricing mechanism's properties differently. Since the ramping constraint reflects the allocation of limited
capacity among energy and ramping services for a resource in the electricity markets\cite{7779055}. When adopting traditional LMP, generators' LMP is influenced by the binding ramping constraints of the last and upcoming periods, but the electricity market operator's merchandise surplus is still composed of the congestion rent. However, in the heat market, heat prices' relationship between periods reflects the heat transfer process, so the heat market operator's merchandise surplus includes $IL_t$ and $IU_t$. The heat nodal balance constraints also couple different nodes in the heat system, so the coupling effect of time and space in heat system needs to be further studied in our future research.
\end{remark}

The total merchandise surplus of system operator over $T$-periods can be calculated as:
\begin{equation}
	\begin{split}
		\Upsilon&=\sum_{i\in \Psi_{L}}\Upsilon_{i}^L-\sum_{i\in \Psi_{S}}\Upsilon_{i}^S=\boldsymbol{-\lambda_{1}^{*T}[C_{2}(T_{0}-T_{a,0})]}\\
		&\boldsymbol{+\sum_{t=1}^T[\mu_{t}^{*T}(T_{sa}-T_{a,t})]}.
	\end{split}
\end{equation}

We can see that the $ IU_{t}$ has been offset between periods. Similarly, the term $ IL_{t}$ has also been offset between periods except the term that reflects the impact of initial temperature- $T_{0}$. So in the total merchandise surplus over $T$-period, there are only two non-negative terms left- congestion rent and the impact from the initial temperature, guaranteeing revenue adequacy for the system operator.

\subsection {Consistency of merchandise surplus under multiple units of temperature}
Note that there are multiple units of temperature, degree centigrade, Fahrenheit, Kelvins, and so on. So we wonder if the merchandise surplus and its decomposition keep the same when we converse the temperature units.

In the conversion of temperature units, the value of temperature $\boldsymbol{T_{t}^U}$ is expressed in the form of a linear transformation: 
\begin{equation}
	\boldsymbol{T_{t}^U=\mathcal{K}T_{t}^*+\mathcal{H}},
\end{equation}
where $\boldsymbol{\mathcal{K}}$ and ${\mathcal{H}}$ are coefficient matrices. After the conversion of units, (9c)'s shadow price $\boldsymbol{\mu_{t}^{*U}}$  has a proportional relationship with $\boldsymbol{\mu_{t}}$:$\boldsymbol{\mu_{t}^{*U}}=\mathcal{K}\boldsymbol{\mu_{t}}$ while $\boldsymbol \lambda_{t}$ keeps the same.

Merchandise surplus at $t$ under any temperature unit $\Upsilon_{t}^{U}$ is calculated as:
\begin{equation}\label{change}
	\begin{split}
		\Upsilon_{t}^{U}
		&=\boldsymbol{-\lambda_{t}^{*T}[C_{2}}/\mathcal{K}\boldsymbol{(T_{t-1}^{U}-T_{a,t}^U)]}\boldsymbol{+\lambda_{t+1}^{*T}[C_{2}}/\mathcal{K}\\
		&\boldsymbol{(T_{t}^{U}-T_{a,t}^U)]+\mu_{t}^{*UT}(T_{sa}^U-T_{a,t}^U)}/\mathcal{K}\\
		&=IL_{t}+IU_{t}+CR_{t}.
	\end{split}
\end{equation}

Comparing (\ref{countt}) and (\ref{change}) we can find that the choice of temperature units does not influence the merchandise surplus and its decomposition.

\vspace{-0.2cm}
\begin{table}[ht]
	\footnotesize
	\centering
	\setlength{\abovecaptionskip}{-0.5pt}
	\setlength{\belowcaptionskip}{0pt}
	\caption{Settlement process of the energy-grade double pricing rule}
	\label{table 4}
	\begin{tabular}{cccc}
		\hline
		Results &Value\\
		\hline
		Generator's LMP ($\$/MWh$) &14.328\\
		Load's LMP ($\$/MWh$)& 14.317\\
		Load's grade price ($\$/^{\circ}C$)&0.0268\\
		Energy payment for generator ($\$$)&30.667\\
		Quality payment from generator ($\$$)&0\\
		Energy payment from load ($\$$)&28.634\\
		Quality payment from load ($\$$)&0.0268*76=2.033\\
		Merchandise surplus ($\$/h$)& 0.00\\
		\hline
	\end{tabular}
\end{table}
\vspace{-0.2cm}

To test the validation of the proposed pricing rule, the same toy case in section III is reviewed. The settlement process is shown in TABLE \ref{table 4}.In the proposed pricing rule, the temperature requirements of NH2 are settled, which guarantees the system operator's revenue adequacy.

\section{Case study}
Detailed data of the testing CHP systems can be found in \cite{test}. All the tests are implemented on a laptop with an Intel Core i5-1035G4 CPU. The optimization problems are solved using the CPLEX solver in MATLAB scripts.

\subsection{ Merchandise surplus' components classification}
\begin{figure}[htbp]
	\vspace{-0.2cm}
	\hspace{0.26cm}
	\includegraphics[width=3.0in]{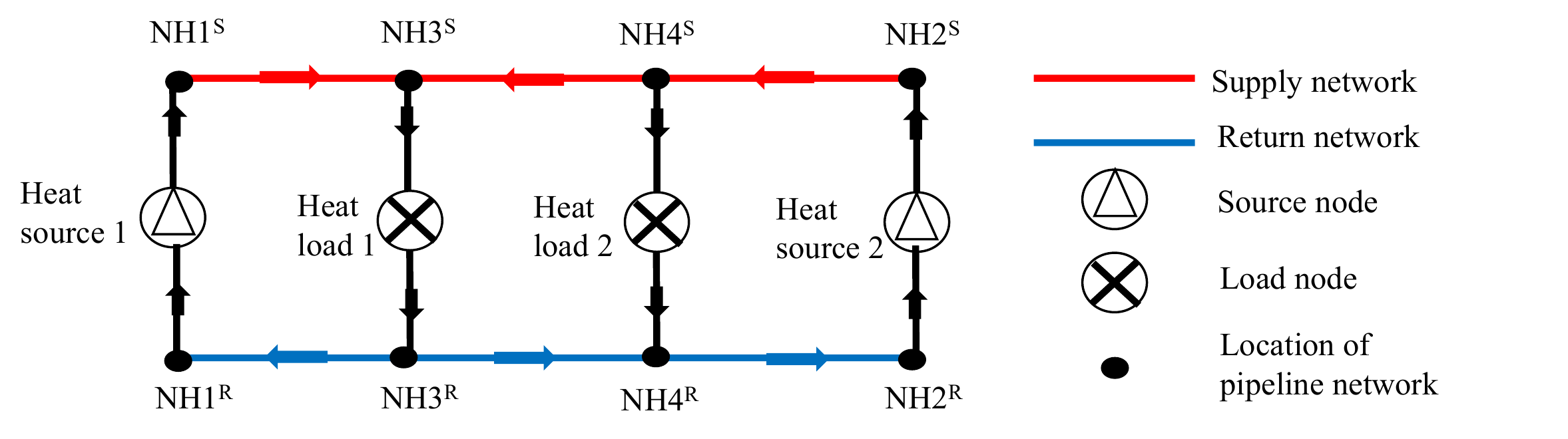}
	\vspace{-0.4cm}
	\caption{four-node heat system.}
	\label{fig 4}
\end{figure}

A 4-node heat system as shown in Fig. \ref{fig 4} is employed to test the validation of merchandise surplus' components classification in the proposed pricing rule. TABLE \ref{table 5} shows merchandise surplus decomposition using different temperature units under the energy-grade double pricing rule. 
	\vspace{-0.3cm}
	
As we discuss in section IV, there are three components in merchandise surplus under the proposed pricing rule - $ CR_{t}$, $ IL_{t}$ and $ IU_{t}$. The congestion rent $ CR_{t}$  is non-negative like the case in the electricity market. Positive $IL_{t}$ reflects that temperatures at 13:00 provide a base for operation at 14:00. Similarly, negative $IU_{t}$ indicates the advance payment of operation state at 14:00 to the operation at 15:00. We also can find that both merchandise surplus and components keep the same under different units of temperatures.
\vspace{-0.3cm}
\begin{table}[ht]
	\footnotesize
	\centering
	\setlength{\abovecaptionskip}{-0.5pt}
	\setlength{\belowcaptionskip}{0pt}
	\caption{Comparison of merchandise surplus components at 14:00}
	\label{table 5}
	\begin{tabular}{cccc}
		\hline
		Units&Centigrade &Fahrenheit& Kelvins \\
		\hline
		$ CR_{t}$ ($\$/h$)  & 7.54 & 7.54&7.54\\
		$ IL_{t}$ ($\$/h$) &184.12 & 184.12&184.12\\
		$ IU_{t}$ ($\$/h$) &-186.42 & -186.42&-186.42\\
		$ IL_{t}+IU_{t}$ ($\$/h$) &-2.30 & -2.30&-2.30\\
		Merchandise surplus ($\$/h$) &5.24&5.24&5.24\\
		\hline
	\end{tabular}
	%\vspace{-0.2cm}
\end{table}
\vspace{-0.6cm}

\subsection{Validation of the energy-grade double pricing rules}
To show validation of the proposed pricing rule in the engineering practical system, a modified DHS of Barry Island \cite{test} is simulated. MSE and MSE-T indicate merchandise surplus of 24 hours in the two rules respectively. In the simulation, MSE=\$ -727.143 and MSE-T=\$0.2470, and their values are not influenced by the choice of temperature units. It shows that under the proposed rule, revenue adequacy holds for heat market operator in the practical system. 

	\vspace{-0.6cm}
\begin{figure}[htbp]
	\vspace{-3.2cm}
	\hspace{0.26cm}
	\includegraphics[width=3.0in]{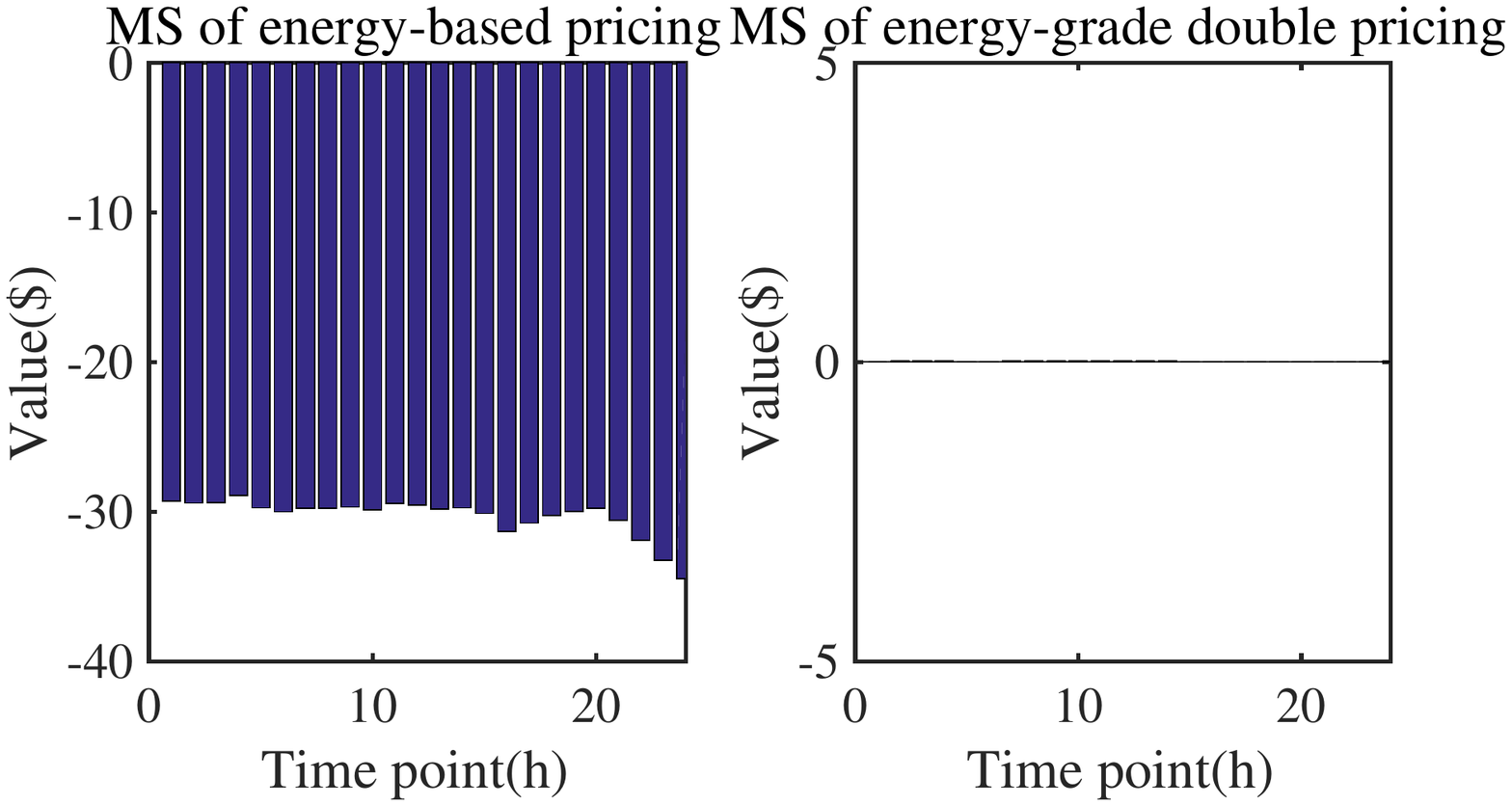}
	\vspace{-3.8cm}
	\caption{Merchandise surplus of energy-based and energy-grade pricing rule.}
	\label{fig 5}
\end{figure}
	\vspace{-0.3cm}
Fig. \ref{fig 5} compares the merchandise surplus at each interval in the two rules. Negative merchandise surplus exists in energy-based pricing rule, because of the high-grade requirements. In the energy-grade double pricing rule, merchandise surplus floats around 0 because of the time-delay effect. 

\section{Conclusion}
Heat is a commodity with a variety of grades. Under the proposed energy-grade double pricing rule, both the heat grade and energy are settled, which guarantees revenue adequacy of system operator, as an improvement from the energy-based pricing rule. It reveals the cost of heat energy and temperature requirements. Case studies show that the explainable merchandise surplus decomposition is valid and revenue adequacy is held in the proposed rule. In addition, different temperature units do not influence the merchandise surplus and its decomposition. 

\bibliographystyle{IEEEtran}
\bibliography{IEEEabrv,ref}

\begin{appendix}
The merchandise surplus of the system operator at time $t$ is calculated as:
\begin{subequations}
	\begin{align}
		\Upsilon_{t}&=\sum_{i\in \Psi_{L}}\Upsilon_{i,t}^L-\sum_{i\in \Psi_{S}}\Upsilon_{i,t}^S\\
		&=\boldsymbol{\mu_{t}^{*T}T_{sa}-\beta_{t}^{*T}T_{Q}-\lambda_{t}^{*T}C_{2}T_{t-1}^*+\lambda_{t+1}^{*T}C_{2}T_{t}^*}\notag\\
		&\boldsymbol{-\lambda_{t}^{*T}R_{t}+\beta_{t}^{*T}(T_{Q}-T_{a,t})}\\
		&=\boldsymbol{\mu_{t}^{*T}(T_{sa}-T_{a,t})-\beta_{t}^{*T}(T_{Q}-T_{a,t})}\notag\\
		&\boldsymbol{-\lambda_{t}^{*T}C_{2}(T_{t-1}^*-T_{a,t})+\lambda_{t+1}^{*T}C_{2}(T_{t}^*-T_{a,t})}\notag\\
		&\boldsymbol{-\lambda_{t}^{*T}A_{a2}G_{c}(D-I)A_{a1}^TT_{a,t}+\beta_{t}^{*T}(T_{Q}-T_{a,t})}\notag\\
		&\boldsymbol{-(\lambda_{t}^{*T}C_{2}+\lambda_{t}^{*T}C_{1})T_{a,t}}\notag\\
		&\boldsymbol{+(\lambda_{t}^{*T}C_{1}+\mu_{t}^{*T}-\beta_{t}^{*T}+\lambda_{t+1}^{*T}C_{2})T_{a,t}}\\
		&\boldsymbol{=\mu_{t}^{*T}(T_{sa}-T_{a,t})-\beta_{t}^{*T}(T_{Q}-T_{a,t})}\notag\\
		&\boldsymbol{-\lambda_{t}^{*T}C_{2}(T_{t-1}^*-T_{a,t})+\lambda_{t+1}^{*T}C_{2}(T_{t}^*-T_{a,t})}\notag\\
		&\boldsymbol{-\lambda_{t}^{*T}A_{a2}G_{c}(D-I)A_{a1}^TT_{a,t}+\beta_{t}^{*T}(T_{Q}-T_{a,t})}\notag\\
		&\boldsymbol{-(-\lambda_{t}^{*T}A_{a2}G_{c}DA_{a1}^{T}+\lambda_{t}^{*T}A_{a1}G_{c}A_{a1}^{T})T_{a,t}}\\
		&\boldsymbol{=\mu_{t}^{*T}(T_{sa}-T_{a,t})-\beta_{t}^{*T}(T_{Q}-T_{a,t})}\notag\\
		&\boldsymbol{-\lambda_{t}^{*T}C_{2}(T_{t-1}^*-T_{a,t})+\lambda_{t+1}^{*T}C_{2}(T_{t}^*-T_{a,t})}\notag\\
		&\boldsymbol{+\beta_{t}^{*T}(T_{Q}-T_{a,t})+\lambda_{t}^{*T}(A_{a2}-A_{a1})G_{c}A_{a1}^TT_{a,t}}\\
		&\boldsymbol{=-\lambda_{t}^{*T}C_{2}(T_{t-1}^*-T_{a,t})+\lambda_{t+1}^{*T}C_{2}(T_{t}^*-T_{a,t})}\notag\\
		&\boldsymbol{+\mu_{t}^{*T}(T_{sa}-T_{a,t})}.
	\end{align}
\end{subequations}			

The transformation from (27a) to (27b)is based on the KKT condition:$\boldsymbol{\frac{\partial L_{CP}}{\partial T_{t}}=0}$ and (9a) as (28) shows:

\begin{subequations}
	\begin{align}
		\boldsymbol{(\frac{\partial L_{CP}}{\partial T_{t}})^TT_{t}^*}&=\boldsymbol{(\lambda_{t}^{*T}C_{1}+\lambda_{t+1}^{*T}C_{2}+\mu_{t}^{*T}-\beta_{t}^{*T})T_{t}^*}\\
		&\boldsymbol{=\lambda_{t}^{*T}(H_{t}-R_t)-\lambda_{t}^{*T}(C_{2}T_{t-1}^*)}\notag\\
		&\boldsymbol{+\lambda_{t+1}^{*T}(C_{2}T_{t}^*)+\mu_{t}^{*T}T_{sa}-\beta_{t}^{*T}T_{Q}=0}\\
		\boldsymbol{\lambda_{t}^{*T}H_{t}}&=\boldsymbol{\sum_{i\in \Psi_{HN}}\lambda_{i,t}^**(G_{i,t}^*-D_{i,t}^*)}.
	\end{align}
\end{subequations}
The transformation from (27c) to (27d) is also based on the KKT condition:$\boldsymbol{\frac{\partial L_{CP}}{\partial T_{t}}=0}$.
Moreover, based on the hydraulic balance of nodes: $\boldsymbol{(A_{a2}-A_{a1})G_{c,t}=0}$ \cite{7243359}, we can get (27f) from (27e).

\end{appendix}
\end{document}